\documentclass[twocolumn,pra,aps,showpacs,superscriptaddress,floatfix,showkeys,nofootinbib,10pt]{revtex4-2}

\usepackage[T1]{fontenc}
\usepackage{amsmath,amsfonts,amssymb,amsthm}
\usepackage{dsfont}
\usepackage[toc,page]{appendix}
\usepackage{bbm}
\usepackage{enumitem}
\usepackage{framed}
\usepackage[labelformat=parens]{subfig}
\usepackage{tabularx}
\usepackage{booktabs}
\usepackage{graphicx}
\usepackage{siunitx}
\usepackage{tikz}
\usetikzlibrary{arrows.meta}

\newcounter{rowcntr}[table]
\renewcommand{\therowcntr}{\thetable.(\alph{rowcntr})}

\newcolumntype{N}{>{\refstepcounter{rowcntr}\therowcntr}c}

\AtBeginEnvironment{tabular}{\setcounter{rowcntr}{0}}

\newcommand{\ket}[1]{ | \, #1 \rangle} \newcommand{\bra}[1]{ \langle #1 \, |}

\newcommand{\be}{\begin{equation}} \newcommand{\ee}{\end{equation}}
\newcommand{\ba}{\begin{aligned}} \newcommand{\ea}{\end{aligned}}

\DeclareMathOperator{\Tr}{Tr}

\DeclareRobustCommand\openone{\leavevmode\hbox{\small1\normalsize\kern-.33em1}}%

\begin{document}

\title{Extensive search of Shannon entropy-based randomness certification protocols}

\author{Robert Okula} \email{rbrt.okula@gmail.com}
\affiliation{Department of Algorithms and System Modeling, Faculty of Electronics, Telecommunications and Informatics, Gda\'{n}sk University of Technology, Poland}
\affiliation{Department of Physics, Stockholm University, 106 91 Stockholm, Sweden} 

\author{Piotr Mironowicz} 
\affiliation{Department of Algorithms and System Modeling, Faculty of Electronics, Telecommunications and Informatics, Gda\'{n}sk University of Technology, Poland}
\affiliation{Center for Theoretical Physics, Polish Academy of Sciences, Aleja Lotników 32/46, 02-668 Warsaw, Poland}

\date{\today}

\begin{abstract}
Quantum technologies offer significant advancements in information processing and communication, notably in the domain of random number generation (RNG). The use of Bell inequalities enables users to certify the randomness of outputs produced by untrusted quantum RNG devices. We present a method for quantitatively analyzing Bell expressions used to certify randomness in quantum systems. Using this method, we conducted a comprehensive analysis on more than half a million Bell expressions involving configurations with four measurement settings for one party and three for the other. We identified five notable examples based on entropy scores under varying levels of white noise. As an extension of these results, we further incorporate the concept of self-testing for boxes (Banacki et al 2022, New J. Phys. 24 083003), enabling a more comprehensive characterization of quantum correlations through the evaluation of $Boxes(\alpha, B)$ and the corresponding measure $Flex(\alpha, B)$. Our results provide a systematic tool for selecting Bell expressions that maximize certifiable randomness under realistic noise levels, directly informing the design and benchmarking of device-independent quantum random number generators. We also demonstrate the practical utility of our certification framework by applying it to experimental data from several quantum random number generators, showing that despite realistic device imperfections, Shannon-entropy-based certification and extraction achieve higher certifiable randomness rates than standard min-entropy methods.
\end{abstract}


\maketitle

\section{Introduction} \label{sec:introduction}
In 1964, John Bell demonstrated \cite{PhysicsPhysiqueFizika.1.195} that the predictions of quantum theory are fundamentally incompatible with the assumption that physical properties are predefined prior to measurement and that a system can only be affected by its local environment. His theorem marks the beginning of quantum information theory by demonstrating how quantum resources can outperform classical ones in generating correlations, inspiring a range of tools for practical communication and processing.

In this theorem, a scenario is considered in which two independent and spatially separated parties, Alice and Bob, together with a third entity acting as a referee, are involved. The referee distributes an entangled quantum state to Alice and Bob in the form of a bipartite entangled system, such as a pair of polarization-entangled photons, with one particle sent to each party. Upon receiving their respective subsystems, Alice and Bob perform local measurements, randomly selecting from a set of observables: ${M^{0,0}, M^{0,1}, \ldots}$ for Alice and ${M^{1,0}, M^{1,1}, \ldots}$ for Bob. For instance, when considering binary observables, we have $M^{p,x} = \Pi_+^{p,x} - \Pi_-^{p,x}= 2 \Pi_+^{p,x} - \openone$. Let us consider a linear combination of the expected values for these measurements, for example the CHSH operator~\cite{PhysRevLett.23.880}:
\begin{equation} \label{eq:chsh}
    G_{CHSH} = C(0,0) + C(0,1) + C(1,0) - C(1,1),
\end{equation}
where $C(x,y)$ is a correlation operator:
\begin{equation} \label{eq:correlator}
\begin{split}
	C(x,y) &= M^{0,x} \otimes M^{1,y} \\
    &= 4 \Pi_+^{0, x} \Pi_+^{1, y} - 2 \Pi_+^{0, x} - 2 \Pi_+^{1, y} + \openone,
\end{split}
\end{equation}
$x$ and $y$ stand for the measurement settings, Alice's projector on results $+1$ ($-1$) is denoted as $\Pi_{+}^{0, x}$ ($\Pi_{-}^{0, x}$), and analogously for Bob $\Pi_{+}^{1, y}$ ($\Pi_{-}^{1, y}$). We can also denote the expectation value of the CHSH operator as: $\langle G_{CHSH} \rangle = \langle C(0,0) \rangle + \langle C(0,1) \rangle + \langle C(1,0) \rangle - \langle C(1,1) \rangle$.

If the measured properties of the objects exist independently of observation (which we call \textit{realism}) and if the choice of the observable by one party cannot influence the results of the other (which we call \textit{locality}), the upper bound of the absolute value of the expectation value of \eqref{eq:chsh} is $|\langle G_{CHSH} \rangle| \leqslant 2$. This limit is referred to as the classical (or local) bound. The expression in the form of \eqref{eq:chsh} is commonly called CHSH and is a member of a group that we call Bell expressions. When presented as an inequality involving its classical bound $\mathfrak{B}$: $|\langle G \rangle | \leqslant \mathfrak{B}$, it is referred to as a Bell inequality.

However, these classical assumptions do not universally hold within the framework of quantum mechanics. In this case, the measurement outcomes of the two parties are fundamentally linked through non-local correlations; the state of the entire system of entangled particles is such that independent local hidden variables cannot describe the results.

When considered independently, the absolute value of the expectation value for each correlation operator satisfies $|\langle C(x,y) \rangle| \leqslant 1$, reflecting the $\pm 1$ eigenvalues of the local measurement observables. However, when we try to maximize the expectation value of the full expression \eqref{eq:chsh} over quantum states and measurements, the value of each individual expectation value becomes $1/\sqrt{2}$, which leads to the quantum upper bound for the expression. This arises because, although each individual observable can attain perfect correlation, the set of four correlations entering the CHSH combination cannot be simultaneously maximized due to the noncommutativity of the underlying quantum operators. The upper bound of a Bell expression for a fully entangled quantum state is called Tsirelson's bound \cite{cirelson_quantum_1980}, and for CHSH it equals $2 \sqrt{2}$. If a Bell expression yields a value exceeding its classical bound, the state is at least partially entangled. Moreover, the degree to which the Bell expression exceeds the classical bound serves as a quantifiable resource related to the entanglement. When the expression reaches Tsirelson’s bound, the state is maximally entangled. Since the derivation of the CHSH expression, a plethora of other Bell expressions have been formulated \cite{PhysRevLett.61.662, PhysRevA.87.020302, PhysRevLett.129.150403, PhysRevA.78.032112}, e.g., for detection of states that are partially separable \cite{Gisin_1998, Seevinck_2002, PhysRevLett.88.230406} or using the property of Frank-Wolfe algorithms for constructing Bell inequalities with a high resistance to noise \cite{Designolle_2023, PhysRevA.96.012113, brierley2017convexseparationconvexoptimization}.

\subsection{Randomness}

It is well established that classical computers, which rely on deterministic algorithms, cannot generate truly random numbers \cite{Vattulainen_1995}. Rather, they generate pseudo-random values that only mimic randomness but are ultimately derived from inherently deterministic processes or computations. Given knowledge of the algorithm and its initial input (or seed) one can reproduce the entire output sequence, even if the output of such an algorithm presents high entropy. The generation of true randomness necessitates access to a fundamentally non-deterministic physical process. The intrinsically probabilistic nature of quantum phenomena serves as a powerful resource for achieving this objective. The validity of the randomness is especially important for the security of quantum cryptography as a whole, including quantum key distribution \cite{Bennett_2014, PhysRevLett.67.661, 8527822, 5204783}.

Bell's theorem enables the certification of randomness and the unpredictability of measurement outcomes in cryptographic protocols \cite{colbeck2011quantumrelativisticprotocolssecure}, by ruling out local hidden-variable models, in which outcomes are predetermined and therefore perfectly predictable given the hidden variables. A Bell-inequality violation constrains the adversary’s guessing probability $P_{guess}$, yielding a rigorous lower bound on certified entropy (e.g., min-entropy) derived solely from observed input-output statistics, without assumptions about the devices’ internal operation, which makes it device-independent \cite{Ac_n_2007} (DI).

This method has been initially presented in \cite{pironio_random_2010}, based upon the violation of the CHSH expression. In \cite{Mironowicz_2013}, a method for analysis of Bell expressions used for the randomness certification has been presented. The authors examine the efficacy of several Bell expressions in the presence of white noise and show which one can certify up to two bits of randomness per measurement, with min-entropy as the measure of randomness. We have expanded this idea further in \cite{okuła2025deviceindependentshannonentropycertification}, where we have developed a method for the randomness analysis and certification that utilizes Shannon entropy.

In this paper, we propose a method for exploring and comparing a wide range of Bell expressions in the context of randomness certification. We then use it to perform an extensive evaluation of the complete set of expressions with unitary coefficients for four measurement settings for Alice and three measurement settings for Bob.

While many Bell inequalities have been proposed and studied analytically, and there are also frameworks for numerical quantitative analysis \cite{Navascués_2008, PhysRevLett.98.010401}, the choice of Bell tests in practical DI QRNG devices is often done without assessment and comparison, including in terms of noise requirements. This absence represents a significant gap in the transition from theoretical protocols to robust physical implementations.

Our results provide a practical decision tool for experimental implementations of DI quantum random number generators, enabling the preselection of Bell expressions that maximize certifiable randomness for a given noise situation and measurement configuration.

The paper is organized as follows: in the latter part of this Section, we present a short review of experimental QRNG methods. In Section \ref{sec:extraction}, we describe the operational details related to the randomness extraction and we present our method for extensive search and comparison of Bell expressions suitable for use as randomness certificates. In Section \ref{sec:results} we present the numerical outcome of the search, including the comparison of several classes of Bell expressions and their certification effectiveness for different levels of noise. We utilize some of the results from our previous work from \cite{okuła2025deviceindependentshannonentropycertification} to develop the search method and we extend the main results of that work by examining the properties of the outcomes obtained there. We also explore criteria for choosing appropriate Bell expressions to ensure robust randomness. In Section \ref{sec:experimentaldatadone}, we demonstrate the practical utility of our certification framework by evaluating its performance using experimental data from several distinct implementations of quantum random-number generators.

\subsection{Experimental QRNG implementations} \label{ss:baselinemethodologies}
To date, numerous DI quantum random number generators (QRNGs) have been proposed \cite{RevModPhys.89.015004}, employing a variety of physical mechanisms \cite{xavier2010practicalrandomnumbergeneration, MONGIA2024129954, Shalm2021, Kwon_2009, Xu:16, Seguinard_2023, Wei:09, PhysRevApplied.22.044041}. The selected experimental datasets cover a range of DI and semi-DI (SDI) randomness generation protocols. The original studies used different information-theoretic methods to estimate min-entropy. Reviewing their assumptions and extraction methods provides the context needed to apply the Shannon-entropy framework.

Early work on DI randomness generation established methods for handling imperfections and providing rigorous certification. In \cite{Pironio2010}, the authors certified the creation of new randomness in an experiment without requiring a detailed model of the devices, which was the first time DI certification was realized. They have used atomic qubits. While the setup achieved near-perfect state detection with an error rate of less than 3\% to close the detection loophole (which arises when imperfect detector efficiency allows the observed sample of entangled particles to be non-representative of the full ensemble, enabling local hidden-variable models to reproduce apparent violations of Bell's Theorem), the probabilistic nature of the entanglement swapping process resulted in a heralded entanglement success probability. Because the same untrusted devices were used repeatedly in a DI setting, the analysis had to account for possible memory effects across sequential measurements. The analysis bounded the min-entropy using SDP to find the minimal value of min-entropy compatible with a given quantum theoretical Bell violation, establishing (together with \cite{colbeck2011quantumrelativisticprotocolssecure}) a framework in which Bell inequality violations are translated into quantitative entropy bounds.

Subsequent photonic implementations developed these ideas under more realistic experimental conditions, representing different models for quantum randomness generation and certification. In \cite{Liu2018}, the authors implemented a detection-loophole-free DI QRNG based on entangled photons. Their analysis does not rely on the i.i.d.\ assumption and explicitly considers security against general quantum adversaries. Randomness is quantified via smooth min-entropy derived from Bell inequality violations, with large-scale Toeplitz matrix hashing used for extraction \cite{Impagliazzo1989Extractor, Frauchiger2013TrueRandomness}, achieving certified generation rates around $0.00114$ bits/run. In \cite{Shen2018}, a similar approach was tried, however, the authors considered a continuous-wave SPDC source and analyzed the trade-off between Bell violation and measurement round duration. In this case, the achievable randomness rate is limited by Poissonian multi-pair emissions and detector timing jitter, with final extraction performed using composable one-shot extractors secure against quantum adversaries.

Beyond fully DI approaches, alternative models relax some assumptions to improve practicality. In \cite{Leone2022}, the authors proposed an SDI protocol based on single-photon momentum–polarization entanglement generated with attenuated lasers. Their approach replaces full DI with a detailed physical model that incorporates detector imperfections, such as dead time and after-pulsing, allowing estimation of the guessing probability and certification of min-entropy under explicit assumptions on the system.

Finally, more recent work has moved toward system-level implementations that integrate certified randomness into broader cryptographic tasks. In \cite{Lin2023}, the authors studied a free-space EPR-based setup subject to significant transmission loss and finite-size effects. Their analysis uses quantum probability estimation to bound the min-entropy and supports the deployment of randomness in a networked setting. In particular, such implementations enable publicly accessible randomness beacons combined with post-quantum cryptographic authentication, illustrating how DI randomness can be incorporated into applications such as zero-knowledge proofs without relying on idealized random oracle assumptions.

We present the comparison of these experimental methods in Table~\ref{tab:qrng_comparison}. Together, they span different physical models of QRNG and therefore constitute a representative benchmark set for evaluating our method against existing approaches.

\begin{table*}[t]
\centering
\caption{Comparison of representative implementations of quantum randomness generation and certification.}
\label{tab:qrng_comparison}
\begin{tabular}{l l l l l}
\toprule
\textbf{Work} & \textbf{Platform} & \textbf{Model} & \textbf{Key assumptions} & \textbf{Extraction method and measure} \\
\midrule

\cite{Pironio2010} 
& Atomic qubits 
& DI 
& No i.i.d.; memory effects allowed
& SDP + martingale min-entropy bounds \\

\cite{Liu2018}
& Entangled photons 
& DI 
& No i.i.d.; quantum adversary
& Smooth min-entropy + Toeplitz hashing \\

\cite{Leone2022} 
& Single-photon entanglement 
& SDI 
& Trusted modeling (detectors, source) 
& Markov model + min-entropy estimation \\

\cite{Lin2023} 
& Free-space EPR links 
& DI
& Finite-size + channel loss
& Quantum probability estimation \\

\cite{Shen2018} 
& CW SPDC photons 
& DI 
& Finite-time analysis 
& One-shot entropy + composable extractors \\

\hline
\end{tabular}
\end{table*}

\section{Randomness extraction and certification} \label{sec:extraction}
In the analysis, we consider a bipartite scenario as the source of randomness, which is the standard for the self-testing and DI QRNG \cite{Mannalatha_2023}. The parties are spatially separated, precluding classical signaling; consequently, they cannot exchange information regarding measurement choices or influence one another's settings, ensuring independent selection. The parties also operate on the binary outcomes, and the number of inputs (measurement settings) in our analysis does not exceed four for the first party (Alice) and three for the second party (Bob). Although our framework works in any general configuration, we utilize this configuration as a practical trade-off – it expands beyond the $(2\times2)$ CHSH scenario while keeping the SDP optimization computationally tractable. However, only some pairs of settings are going to maximize randomness.

The parties share a two-qubit state subject to isotropic depolarizing (white) noise, provided by a third party or prepared prior to the protocol:\begin{equation} \label{eq:werner}
\rho = p \frac{\openone}{4} + (1 - p) \ket{\psi} \bra{\psi},
\end{equation}
where $\bra{\psi}$ is a state that achieves the maximal quantum violation of the corresponding Bell inequality and $p$ is the noise level.

Based on \eqref{eq:correlator}, we can then define the general form of the expression with unitary coefficients:
\begin{equation} \label{eq:galpha}
	G_{\alpha} = \sum_{i \leq N,j \leq M} \alpha_{i,j} C(i,j),
\end{equation}
where $\alpha_{i,j} \in \{-1,0,1\}$, $N$ and $M$ denote the number of measurement settings for Alice and Bob, respectively, with $i$ and $j$ labeling their corresponding settings. The expression in this form can be described by its coefficient matrix:
\begin{equation} \label{eq:alpha}
\alpha = \begin{bmatrix} 
    \alpha_{1,1} & \alpha_{1,2} & \dots \\
    \alpha_{2,1} & \ddots & \\
    \vdots &        & \alpha_{N,M} 
    \end{bmatrix}.
\end{equation}
However, in the entropy-based analysis, it is more convenient to work directly with the underlying probability distributions. Let Alice’s measurement outcome be $a$ and Bob’s be $b$, each associated with projectors $\Pi^{0,x}_a$ and $\Pi^{1,y}_b$, respectively. The corresponding joint probability distribution is then
\begin{equation} \label{eq:probdisfullt}
	P(a,b|x,y) = \Tr \left[\rho \left(\Pi^{0,x}_a \otimes \Pi^{1,y}_b \right) \right].
\end{equation}
In the binary case, outcomes satisfy $a, b \in \{-1,1\}$. We define a correlator in a form of:
\begin{equation} \label{eq:pcorrelator}
\begin{split}
	\bar{C}(x,y,P) &= P(a=-1,b=-1|x,y)\\
    &- P(a=-1,b=1|x,y)\\
	&- P(a=1,b=-1|x,y)\\
    &+ P(a=1,b=1|x,y),
\end{split}
\end{equation}
By combining this definition with \eqref{eq:probdisfullt}, we obtain the relation between \eqref{eq:correlator} and \eqref{eq:pcorrelator}, which is
\begin{equation}
	 \bar{C}(x,y,P) = \Tr(\rho C(x,y)).
\end{equation}
Given that, the Bell expression with unitary coefficients defined as a function of correlators, instead of correlation operators, takes the form:
\begin{equation} \label{eq:galphap}
	\bar{G}_{\alpha}(P) = \sum_{i \leq N,j \leq M} \alpha_{i,j} \bar{C}(i,j,P).
\end{equation}
To employ this expression as an inequality-based randomness certificate, the user must also specify Tsirelson’s bound $B$ and the pair of measurement spot-settings $(x_0,y_0)$, which are often chosen to maximize the certified randomness. If the expression is defined as a linear combination of correlators $\bar{C}(x,y,P)$, the value of the expression for a state with noise level $p$ and quantum bound $B$ is given by $(1-p)B$. Consequently, a single protocol $R$ can be represented as a set: 
\begin{equation} \label{eq:protocol}
	R = \{ \alpha, B, x_0, y_0 \}.
\end{equation}

\subsection{Entropy as the measure of randomness}
Information entropy has been introduced by Claude Shannon in \cite{shannon}. It is the first and most fundamental measure of information, as it quantifies the unpredictability of the outcome of an operation, based on the probability results. The development of this measure marked the establishment of the fundamental principles of information theory, laying the groundwork for reliable communication over noisy channels and various other applications. It is defined as follows:
\begin{equation}
	H(P) = - \sum_{i=1}^n p_i \log_2 p_i,
\end{equation}
where $P = (p_1, p_2, \ldots, p_n)$ is a probability distribution. It reaches the maximal value when all outcomes are equally likely, viz. the probability of obtaining any results of the operation equals $p_i = \frac{1}{n}$ for $n$ allowed outcomes. This form of entropy quantifies the average uncertainty associated with an outcome, and it also provides a lower bound on the expected number of bits required to encode the information efficiently \cite{10.5555/971143}.

Another type of entropy, more often used in the context of randomness measurement, is called min-entropy. It is defined as:
\begin{equation}
	H_{\infty}(P) = - \log_2 (\max_i (p_i)).
\end{equation}
This form of entropy reflects the unpredictability of the most probable outcome of an event. It is therefore directly linked to the probability of correctly guessing the value of a variable when an observer (e.g., an eavesdropper) guesses the most likely result \cite{10.1007/978-3-319-31875-2_7}. In other words, it represents the minimum possible entropy value (hence the name) and serves as a lower bound for the Shannon entropy, with both measures coinciding in the case of a uniform probability distribution.

Min-entropy is widely used as the metric of the randomness of generated random numbers in cryptographic applications, and it is often used as a measure in randomness certification \cite{pironio_random_2010, Mironowicz_2013, Um2013, Berta_2014, PhysRevA.90.052327, Passaro_2015, Seguinard_2023}. Its relevance arises from the fact that randomness extractors employ min-entropy to characterize the unpredictability of weak random sources, thereby ensuring that the extracted output is statistically close to a uniform distribution only when the source exhibits at least the corresponding amount of entropy.

However, in \cite{okuła2025deviceindependentshannonentropycertification}, we have investigated the topic of randomness certification and randomness estimation using Shannon entropy. The longer the stream of random bits is produced, the better Shannon entropy reflects the average information across many independent samples, as min-entropy considers only the single worst-case outcome.

The practical usability of Shannon entropy in the extraction of quantum randomness is underlined by the existing experimental results and the implementation in real hardware. For example, in \cite{Gabrieletal2010}, Shannon entropy serves as the quantitative foundation for distinguishing irreducible quantum noise from classical noise to ensure unique randomness. The authors utilized Shannon entropy because it provides a direct mathematical framework to quantify the average information content in the Gaussian distribution of vacuum states. They have partitioned the total entropy into distinct quantum and classical components. By calculating the total entropy and subtracting the estimated entropy of classical effects like electronic noise and local oscillator imbalances, they could determine the precise amount of pure quantum-mechanical entropy available. In their setup, they found that even with classical noise contributing 1.75 bits, a significant 3.25 bits of unique quantum entropy remained per sample, justifying a high-speed extraction rate that a more restrictive min-entropy calculation might have unnecessarily penalized.

\subsection{Bounding the probabilities}
Expressions \eqref{eq:pcorrelator} and \eqref{eq:galphap} show, that to calculate the value of the Bell expression, the probabilities of the quantum measurement results for the given measurement settings should be used. In DI randomness certification, an adversary’s optimal guessing probability $P_{\text{guess}}$ is obtained by maximizing the adversary's prediction over all quantum states and measurement strategies compatible with the observed Bell violation. The worst-case analysis requires determining the quantum bounds on outcome probabilities consistent with the observed correlation.

This step utilizes the NPA \cite{Navascués_2008, PhysRevLett.98.010401} method as a semidefinite programming (SDP) technique to obtain the probability bounds, as a form of relaxation. This method provides an infinite hierarchy of semidefinite constraints that any set of quantum correlations must satisfy. Each level of the hierarchy can be formulated as an SDP problem, allowing for efficient numerical optimization. The method analyzes probability distributions arising from projective measurements performed by two parties \eqref{eq:probdisfullt}. The NPA framework is particularly powerful for the numerical characterization of quantum correlations and quantum randomness, making it an essential tool for investigating complex quantum systems and fundamental quantum phenomena.

The most basic utilization of this technique, presented in \cite{Navascués_2008}, is the bounding of the maximal quantum violation, which we denote a $B$. To obtain this value for a Bell expression $\bar{G}$, we need to solve an SDP maximization problem:
\begin{equation}
\label{eq:npalinear2}
\begin{aligned}
	B = \max_{P\in Q_i} \bar{G}_{\alpha}(P).
\end{aligned}
\end{equation}
Maximizing this over each level of NPA hierarchy yields an SDP with solution $\bar{G}_{\alpha}(P)_{Q_i}$ (where $Q_i$ is the set of quantum behaviors for the $i$-level of NPA hierarchy), which serves as an upper bound to the true quantum value $\bar{G}_{\alpha}(P)_{Q}$. Increasing the level systematically tightens the bound (or keeps it unchanged), and the sequence converges to the exact quantum limit in the infinite limit. 

For any Bell inequality-based protocol $R$, we can establish a set of linear expressions $\{L_i\}_i$:
\begin{equation}
	L_i(P, x_0, y_0) = \sum_{a,b} c^{(abx_0y_0)}_i P(a,b|x_0,y_0),
\end{equation}
where $c_i^{(abx_0y_0)}$ is a constant value, and then solve the SDP optimization problem defined as:
\begin{equation}
\label{eq:npalinear}
\begin{aligned}
\max_P \quad & L_i(P, x_0, y_0)\\
	\textrm{subject to} \quad & \bar{G}_{\alpha}(P) - (1-p)B = 0, \\
\end{aligned}
\end{equation}
where $\bar{G}_{\alpha}(P)$ is the value of the Bell expression for an optimized probability distribution and $p$ is the state \eqref{eq:werner} noise level. By solving this optimization problem, we establish the upper and lower bounds for expressions $L_i(P, x_0, y_0)$.

The noise levels $p$ must be carefully selected, since each requires an independent SDP optimization. Selected noise levels should allow for a consistent comparison of the entropy function across protocols and reveal which protocol is most robust to noise. In accordance with the findings in \cite{Mironowicz_2013, okuła2025deviceindependentshannonentropycertification}, the noise rules out reliable entanglement-based randomness generation for $p > 0.2$ (because the lower bounds on Shannon and min-entropy drop sharply across all tested Bell inequalities, meaning the device’s outputs can no longer be certified as reliably random). Therefore, for this analysis, we put forward three levels of noise: $p = 1\mathrm{e}{-6}$, $p = 0.1$, and $p = 0.2$.

\subsection{Calculating entropy}
Typically, the estimation of entropy for the bounded quantum behaviors is based on the NPA optimization of guessing probability \cite{Mironowicz_2013}, but for Shannon entropy, the accurate estimation requires nonlinear optimization over the feasible set of probability distributions consistent with the observed statistics \cite{okuła2025deviceindependentshannonentropycertification}. Furthermore, it demands a high number of iterations to avoid local entropy minima. This limitation motivates the search for analytic or semi-analytic methods that can provide accurate entropy bounds without resorting to full nonlinear optimization. In \cite{okuła2025deviceindependentshannonentropycertification}, two complementary approaches towards this goal had been presented.

First, a rigorous analytical lower bound on the Shannon entropy is provided, albeit only on the assumption that limited information about the distribution is available – specifically, that each element of the normalized probability distribution satisfies $p_i \leqslant u_P$. 
\begin{equation}
    \label{eq:prop1}
    H(P) \geqslant -u_P \log_2(u_P) - (1-u_P) \log_2(1-u_P).
\end{equation}
This inequality results from the concavity of the Shannon entropy and corresponds to the minimal-entropy configuration. Intuitively, the bound assumes that all remaining probability mass is concentrated on one complementary outcome, thereby yielding the smallest admissible entropy. Importantly, \eqref{eq:prop1} is constraint-agnostic – it does not depend on any specific Bell-inequality structure or DI scenario – and can be applied to arbitrary distributions that satisfy normalization.

In contrast, \cite{okuła2025deviceindependentshannonentropycertification} also provides an empirical ansatz characterization of the entropy-maximizing distribution under multiple linear bounds derived from DI certificates, which utilizes the lower and upper bounds obtained using \eqref{eq:npalinear}, for a set of linear expressions, where: 
\begin{equation}
\label{eq:linexpjawnie}
c^{(-1,-1,x,y)}_1, c^{(-1,1,x,y)}_2, c^{(1,-1,x,y)}_3, c^{(1,1,x,y)}_4 = 1,
\end{equation}
and the remaining constants $c^{(abxy)}_i = 0$ (in other words, the bounded linear expressions are the ones that correspond to single probabilities of outcomes from quantum behavior). We establish the lower and upper bounds for linear expressions \eqref{eq:linexpjawnie}, which we denote as $l_{ab}$ and $u_{ab}$, respectively. The probability distribution that maximizes the value of Shannon entropy in this scenario was observed to take the form (up to permutation of outcomes) is then:
\begin{equation} \label{eq:hipo}
    P_H = [l_{--}; l_{-+}; u_{+-}; u_{++}],
\end{equation}
where $l_{--}$ is the lower bound for the results $a=-1, b=-1$, $l_{-+}$ is the lower bound for the results $a = -1, b=1$ and $u_{+-}$ is the upper bound for the result $a = 1, b = -1$ and $u_{++} = 1 - l_{--} - l_{-+} - u_{+-}$. While this form has not been proven for all Bell-inequality families, numerical optimization confirms its validity to at least six decimal places for all cases studied in \cite{okuła2025deviceindependentshannonentropycertification}. 

\section{Quantitative Analysis of Bell Expressions} \label{sec:results}
We present an extensive numerical investigation and comparative analysis of Bell certificates for configurations with four measurement settings on one side and three on the other, conducted using the proposed quantitative procedure. Section \ref{ssec:generalanalysis} shows the global results and the distribution of the Bell certificates, based on the certifiable Shannon entropy. In Section \ref{ssec:selectedprotocols}, we focus on identifying Bell certificates with particularly favorable properties for DI randomness certification by examining representative protocols, comparing their entropy behavior under varying noise levels, and assessing the accuracy of entropy estimates obtained from the analytical distribution~\eqref{eq:hipo} relative to nonlinear optimization results. Finally, Section \ref{ssec:boxestesting} investigates the self-testing properties of the identified Bell certificates within the framework of self-testing for boxes introduced in~\cite{banacki_hybrid_2022}.

\subsection{General analysis}
\label{ssec:generalanalysis}
Building on the definition of the coefficient matrix in~\eqref{eq:alpha} and the general form of the Bell operator in~\eqref{eq:galpha}, we first quantify the size of the search space considered in our analysis, corresponding to all possible Bell expressions constructed from discrete unit coefficients.

For $N = 4$ measurement settings on Alice’s side and $M = 3$ on Bob’s, the total number of distinct expressions with coefficients $\alpha_{ij} \in \{-1, 0, 1\}$ is $3^{NM} - 1 = 531{,}440$, excluding the trivial zero matrix. With $N \cdot M = 12$ measurement spot-setting combinations $(x_0,y_0)$, this yields $12 \cdot (3^{NM} - 1) = 6{,}377{,}280$ distinct certification protocols $R$\footnote{In principle, if a certificate used with the spot-setting pair $(x_0, y_0)$ includes a non-zero correlator $\bar{C}(x_0,y_0)$, this correlator should be omitted from the Bell expression. During the randomness certification, the goal is to maximize the uncertainty of the results and thus maximize randomness. However, a correlator that is measured using observables that correspond to its $(x_0, y_0)$ settings will always maximize the correlation between two parties. Therefore, such a correlator should be removed from the Bell expression when deployed.}. For each protocol, we employed \eqref{eq:npalinear2} to compute Tsirelson's bound and, using the optimization problem defined in \eqref{eq:npalinear}, performed the probability bounding. These bounds were then used to estimate the corresponding measure of randomness, quantified using Shannon entropy. This procedure has been utilized for three different white noise values: $p = 1\mathrm{e}{-6}$, $p = 0.1$, and $p = 0.2$, as defined in \eqref{eq:werner}.

\begin{figure}[htbp]
	\includegraphics[width=\linewidth]{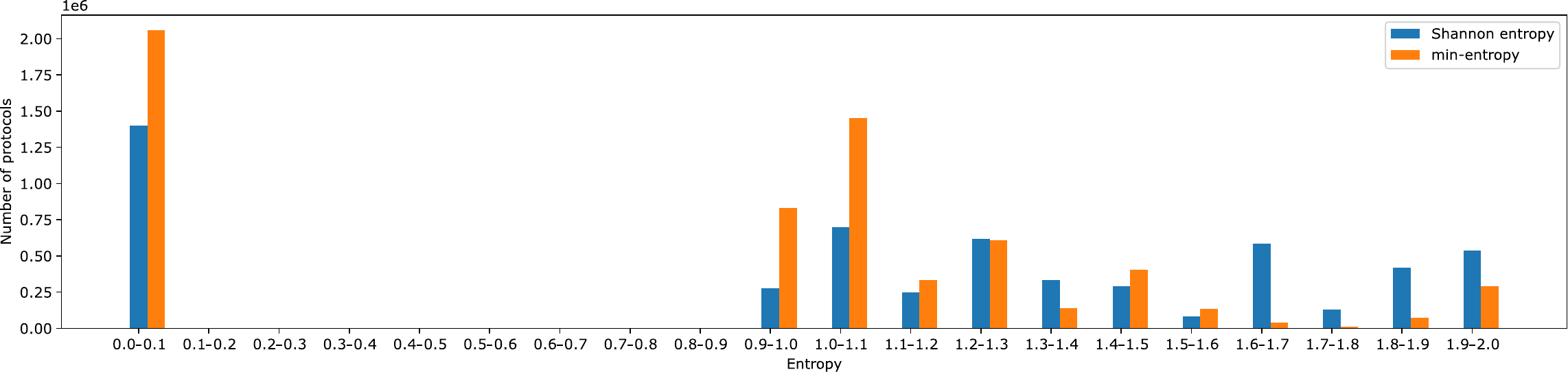}
	\caption{(color online) A distribution of the Bell certificates, as definied by \eqref{eq:protocol}. The certificates are classified based on the value of the measure of randomness (Shannon entropy) that the certificate yields. The histograms are presented for the white noise value $p = 1\mathrm{e}{-6}$, for both Shannon and min-entropy.} 
	\label{fig:HistogramShannon}
\end{figure}

The distribution of Bell certificates, grouped by the value of their associated randomness measure (Shannon entropy), for low noise ($p=1\mathrm{e}{-6}$), is shown in Figure \ref{fig:HistogramShannon}.

Approximately 1.4 million certificates exhibit zero entropy, indicating that they cannot certify any randomness and therefore correspond to Bell values that do not exceed the local bound. Hence, they are not genuine Bell expressions. Approximately 1.4 million certificates exhibit zero entropy. This indicates that for the state family and measurement configurations analyzed here, the expected values do not exceed the classical local bound and thus certify no randomness. Among the remaining certificates, none certifies less than one bit of randomness. Additionally, at $p = 0.2$, the maximum observed entropy is $H_{\text{max}}^{p = 0.2}(P_H) = 1.0724$, while for $p = 0.1$, it increases to $H_{\text{max}}^{p = 0.1}(P_H) = 1.4773$. We have also compared these results with the distribution for min-entropy of the Bell certificates. In this case, more than 2 million certificates exhibit zero entropy.

\subsection{Selected protocols}
\label{ssec:selectedprotocols}
The primary objective of this investigation is to introduce new Bell certificates for use in quantum DI randomness certification.  We explored several protocols exhibiting distinct characteristics in the computed data points. Specifically, we analyzed: protocols that maximize $H^{p = 0.2}(P_H)$ among those satisfying $H^{p = 1\mathrm{e}{-6}}(P_H) \approx 2.0$~\ref{eq:bestc}; protocols that maximize $H^{p = 0.1}(P_H)$ under the same condition, $H^{p = 1\mathrm{e}{-6}}(P_H) \approx 2.0$~\ref{eq:bestf}; protocols exhibiting the smoothest behavior (i.e., minimal variation in the entropy function across different noise levels) at various values of $H^{p = 1\mathrm{e}{-6}}(P_H)$~\ref{eq:bestd},~\ref{eq:beste}; and finally the protocol in~\ref{eq:besta}, where the constraint from~\ref{eq:bestc} was relaxed to $H^{p = 1\mathrm{e}{-6}}(P_H) > 1.7$ instead of $H^{p = 1\mathrm{e}{-6}}(P_H) \approx 2.0$. All of the chosen protocols are presented in Table \ref{eq:best}.

\onecolumngrid

\twocolumngrid

\begin{table}[]
\caption{Protocols chosen for the further analysis, in the form of \eqref{eq:protocol}. All spot-settings equal $(x_0,y_0) = (0,0)$.} \label{eq:best}
\begin{tabular}{|N|l|l|l|l|}
\hline
\multicolumn{1}{|c|}{\textnumero} & \multicolumn{1}{c|}{$\alpha$}    & \multicolumn{1}{c|}{$B$} \\ \hline
\label{eq:besta} & $(0,1,0; -1,-1,0; -1,-1,0; -1,1,0)$ & 5.472136 \\ \hline
\label{eq:bestc} & $(-1,-1,-1; 1,-1,0; 0,0,0; -1,1,1)$ & 6.146067 \\ \hline
\label{eq:bestd} & $(-1,-1,1; -1,-1,0; 1,0,1; 1,0,-1)$ & 6.569963 \\ \hline
\label{eq:beste} & $(1,-1,0; 1,0,-1; 0,0,0; 0,-1,1)$   & 5.196152 \\ \hline
\label{eq:bestf} & $(0,1,-1; 1,1,1; -1,0,-1; 1,1,-1)$  & 7.534802 \\ \hline
\end{tabular}
\end{table}

\begin{figure}[!ht]
    \centering
    \subfloat[\centering Shannon entropy based on \eqref{eq:hipo}]{{\includegraphics[width=\linewidth]{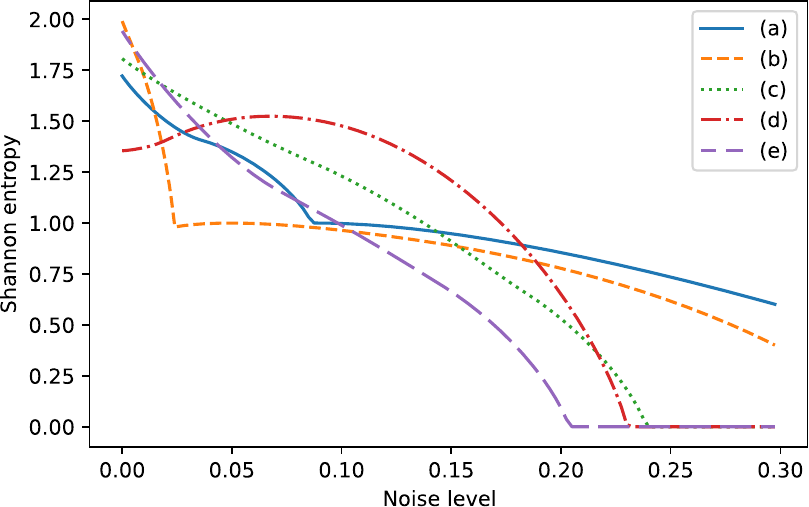} \label{fig:hipo} }}
    \qquad
    \subfloat[\centering Certification of Shannon entropy]{{\includegraphics[width=\linewidth]{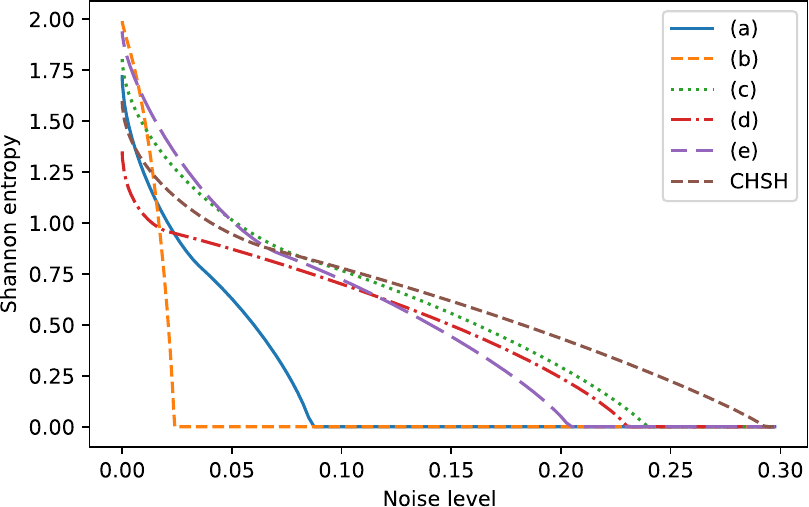} \label{fig:shannon} }}
    \qquad
    \subfloat[\centering Certification of min-entropy]{{\includegraphics[width=\linewidth]{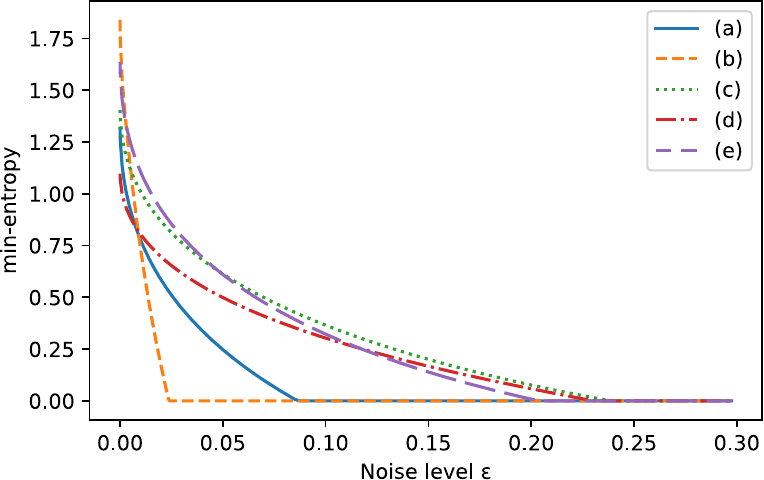} \label{fig:min} }}
    \caption{(color online) Certification of randomness using the Bell inequalities defined in Table \ref{eq:best}, for two measures of randomness. The values are obtained using the standard certification approach, which relies on the nonlinear optimization, constrained by the NPA-obtained bounds, as opposed to \eqref{eq:hipo}.}
    \label{fig:best}%
\end{figure}

For the selected protocols, we computed the Shannon entropy using the distribution defined in~\eqref{eq:hipo} across the range $p \in (1\mathrm{e}{-6}, 0.3)$, with the results presented in Figure~\ref{fig:hipo}. To investigate the issue further, we have also calculated the Shannon Entropy (Figure~\ref{fig:shannon}) using the non-linear optimization protocol presented in \cite{okuła2025deviceindependentshannonentropycertification}. 

\begin{figure}[htbp]
	\includegraphics[width=\linewidth]{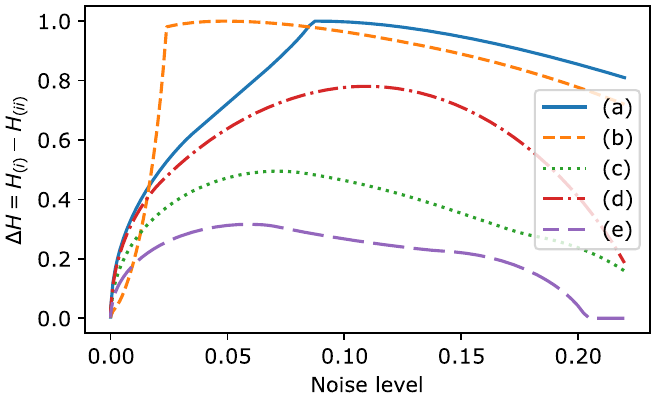}
	\caption{(color online) Difference between the Shannon entropy values obtained from \eqref{eq:hipo} and those derived from nonlinear optimization, illustrating the overestimation of entropy by \eqref{eq:hipo} at higher noise levels.}
	\label{fig:difference}
\end{figure}

The results showed that the value of entropy based on \eqref{eq:hipo} for every analyzed certificate is higher than the entropy values for optimized quantum behavior. The results have even presented that \eqref{eq:hipo} might even lead to the entropy estimation that is higher for higher values of white noise (as has been shown for \ref{eq:beste}). This suggests that the entropy obtained based on the distribution \eqref{eq:hipo} could be at most considered an upper bound. These results are underlined by the difference between the Shannon entropy values for these two distinct quantum behavior establishment methods, that is presented on Figure \ref{fig:difference}. It is noticeable that for the noise level $p \approx 1\mathrm{e}{-6}$, quantum behavior \eqref{eq:hipo} provides the exact Shannon entropy value. Additionally, it can be noticed that for every Bell expression, the difference peaks for $p \in (0.05, 0.1)$, and then decreases, reaching the minimum value only for entropy that is equal to zero.

These results corroborate the assumption that the Shannon entropy based on the probability distribution \eqref{eq:hipo} can be considered an upper bound for the Shannon entropy obtained using nonlinear optimization.

We have also checked if any certificate from the full set that maximizes the entropy for the highest level of noise can achieve two bits of randomness for the $p = 1\mathrm{e}{-6}$ and the best certificate with this assumption cannot exceed $H^{p = 1\mathrm{e}{-6}}(P_H) = 1.6$, which is equivalent to the standard CHSH. The comparison with CHSH for the inequalities defined in Table \ref{eq:best} has also been presented in Figure \ref{fig:shannon}.

Additionally, we have computed the min-entropy using the method that utilizes non-linear optimization; it is visualized in Figure~\ref{fig:min}. The maximum amount of entropy certified by the protocols from Table \ref{eq:best} does not exceed $H^{p = 1\mathrm{e}{-6}}_{max}(P_{opt}) = 1.8396$, which is achieved for \ref{eq:bestc}.

\subsection{Boxes self-testing}
\label{ssec:boxestesting}
In \cite{banacki_hybrid_2022}, an additional form of self-testing, named \textit{self-testing for boxes}, has been introduced. For a given protocol $R = \{ \alpha, B, x_0, y_0 \}$ and Bell expression $\bar{G}_{\alpha}(P)$, let $Boxes(\alpha, B)$ denote the set of all boxes for the Bell inequality with expression $\bar{G}_{\alpha}(P)$ and quantum bound $B$. In this context, \emph{boxes} refer to sets of probability distributions $\{ P(a,b|x,y) : a,b \in \{ -1, 1 \}, \; x \in \{ 1,2,3,4 \}, \; y \in \{ 1,2,3 \} \}$, for which the value of the operator defined by the Bell expression, denoted by  $\bar{G}_{\alpha}(P)$, attains at least the Tsirelson bound $B$. For a given Bell expression $\bar{G}_{\alpha}(P)$, a corresponding quantity $Flex(\alpha, B)$ is introduced~\cite{banacki_hybrid_2022}:
\begin{equation}
	Flex(\alpha, B) = \frac{1}{NM} \sum_{a,b,x,y} \max_{p \in Boxes(\alpha, B)} p(a,b|x,y).
\end{equation}
Given that, the operator defined by Bell's expression $\bar{G}_{\alpha}(P)$ is self-tested for boxes iff the set $Boxes(\alpha, B)$ has $Flex(\alpha, B) = 1$, which means that the box attaining Tsirelson's bound $B$ is unique. For every certificate, we have tested the flex value for the probability distribution based on the NPA obtained upper bounds for the outcomes. There are 12 probability distributions in this form, one for every pair of settings.

We have verified if the certificates listed in Table \ref{eq:best} are self-tested for boxes. We have observed two protocols, \ref{eq:bestd} and \ref{eq:bestf}, which presented (for $p = 1\mathrm{e}{-6}$) the flex that equals $1.0$, whereas for the other protocols from the list it falls within the vicinity of 1.3.

The lowest flex for the $p = 0.1$ equals $2.4$ and for $p = 0.2$ it equals $2.8$. The normalized flex that is at least $2.0$ means that there are at least two boxes for which the Bell equation yields the $(1-p)B$ value. This degenracy, however, leads to the conclusion, that there are infinitely many boxes that can produce the specific Bell equation value, as for two boxes $Box(\alpha, (1-p)B)_1$ and $Box(\alpha, (1-p)B)_2$, we can define a third box $Box(\alpha, (1-p)B)_3 = \beta_1 Box(\alpha, (1-p)B)_1 + \beta_2 Box(\alpha, (1-p)B)_2$, where $0 > \beta_1, \beta_2 > 1$ and $\beta_1 + \beta_2 = 1$. Thus, there are no certificates that are box-certifiable for $p = 0.1$ or $p = 0.2$.

\begin{table}[htbp]
\caption{Comparison between different certificates presented in Table \ref{eq:best}. Only two of the presented protocols show flex that equals exactly 1, i.e., when the box attaining Tsirelson’s bound is unique.}
\label{tab:boxes}
\begin{tabular}{|l|l|l|}
\hline
Bell expression   & Flex for $p = 1\mathrm{e}{-6}$ & Flex for $p = 0.1$  \\ \hline
\ref{eq:besta}  & 1.338003               & 3.325855            \\ \hline
\ref{eq:bestc}  & 1.255119               & 3.058927            \\ \hline
\ref{eq:bestd}  & 1.003841               & 2.648911            \\ \hline
\ref{eq:beste}  & 1.253785               & 2.829320            \\ \hline
\ref{eq:bestf}  & 1.003749               & 2.663294            \\ \hline
\end{tabular}
\end{table}

\section{Performance Evaluation Using Experimental Data} \label{sec:experimentaldatadone}
To demonstrate the practical utility of our certification framework, we evaluate its performance using experimental data from several distinct implementations of quantum random number generators. Physical devices inevitably introduce imperfections, such as statistical noise and detection inefficiencies, that limit ideal performance. By applying the analytical bounds and extraction methodologies developed in our work to these datasets, we compare our approach with representative experimental implementations of quantum random number generation discussed in Section \ref{ss:baselinemethodologies}. Our goal is not only to contrast performance metrics such as randomness generation rate or efficiency, but also to examine the underlying certification frameworks.

First, we apply our method to the results of \cite{Pironio2010}. In that experiment, $n = 3016$ heralded entanglement events produced a CHSH violation of $B = 2.414$. The original framework certified private randomness by lower-bounding the min-entropy of the outputs, conditioned on randomly chosen measurement settings, while accounting for statistical fluctuations via the Azuma–Hoeffding inequality \cite{Azuma1967} at 99\% confidence. This finite-size analysis yielded 42 certified random bits. In contrast, the corresponding penalty-free asymptotic limit (i.e., assuming infinitely many trials and no statistical corrections) gives a theoretical maximum of 1088 bits.

We contrast these baselines with the yields obtained by applying our Shannon entropy-based methodology introduced in \cite{okuła2025deviceindependentshannonentropycertification}. To provide a direct comparison that maintains the original finite-size statistical penalties, we first utilize the analytical Shannon lower bound. We utilize the penalized maximum guessing probability derived from the experimental data in \cite{Pironio2010} into the Shannon entropy formula, and this way, we certify 230 secure bits, demonstrating the immediate yield advantage of measuring average unpredictability over worst-case min-entropy. Furthermore, we evaluate the maximum certified limit using the full joint Shannon optimization procedure over the complete, multi-parameter outcome distribution. 

However, although Shannon entropy certification characterizes the asymptotic randomness rate, finite-key security requires a lower bound on the smooth min-entropy conditioned on adversarial quantum side information. Using the quantum AEP \cite{Tomamichel_2009}, the smooth min-entropy is bounded by the Shannon entropy up to finite-size corrections scaling with the number of trials. The Leftover Hash Lemma \cite{Tomamichel2011} then determines the extractable secret randomness after privacy amplification, which guarantees composable security against quantum adversaries. We compare all of the results in Table \ref{tab:randomness_comparison}.

\begin{table}[bth]
\centering
\begin{tabular}{l c}
\toprule
\textbf{Framework} & \textbf{Random bits} \\
\midrule
\cite{Pironio2010} full extraction (Min-Entropy) & 42   \\
\cite{okuła2025deviceindependentshannonentropycertification} analy. lower bound (Shannon) & 230  \\
\cite{Pironio2010} asymptotic limit (Min-Entropy) & 1088 \\
\cite{okuła2025deviceindependentshannonentropycertification} extracted limit (Shannon) & 2145 \\
\bottomrule
\end{tabular}
\caption{Comparison of certified private random bits extracted from the $n = 3016$ experimental trials reported in \cite{Pironio2010} under various theoretical frameworks.}
\label{tab:randomness_comparison}
\end{table}

Continuing, we evaluate our certification method in the \cite{Liu2018} experimental setting. The extraction method presented in this paper was based on a smoothed min-entropy certification procedure based on the Entropy Accumulation Theorem (EAT) \cite{Dupuis_2020}, which does not assume i.i.d. trials. In this approach, randomness is accumulated securely over finite experimental runs. Optimization methods such as basin-hopping are applied to the EAT bounds to identify the tangent point that maximizes net randomness while accounting for finite-size statistical fluctuations. Conditioned on successful CHSH violation during random spot-checking, the raw data is processed using a Toeplitz-matrix hashing extractor to generate a uniform random bit string.

As a contrast, we utilize these measurement data to obtain the certifiable Shannon entropy for this experiment, as described by our certification method presented in \cite{okuła2025deviceindependentshannonentropycertification}. This way, we have presented the possibility of certifying $6.83 \cdot 10^{-2}$ bits/run. Additionally, we calculate the upper bound of the extractable bits per single run presented in \cite{10.1007/978-3-319-31875-2_7}, which equals approximately $6.81 \cdot 10^{-2}$ bits/run. 

\begin{figure}[htbp]
	\includegraphics[width=\linewidth]{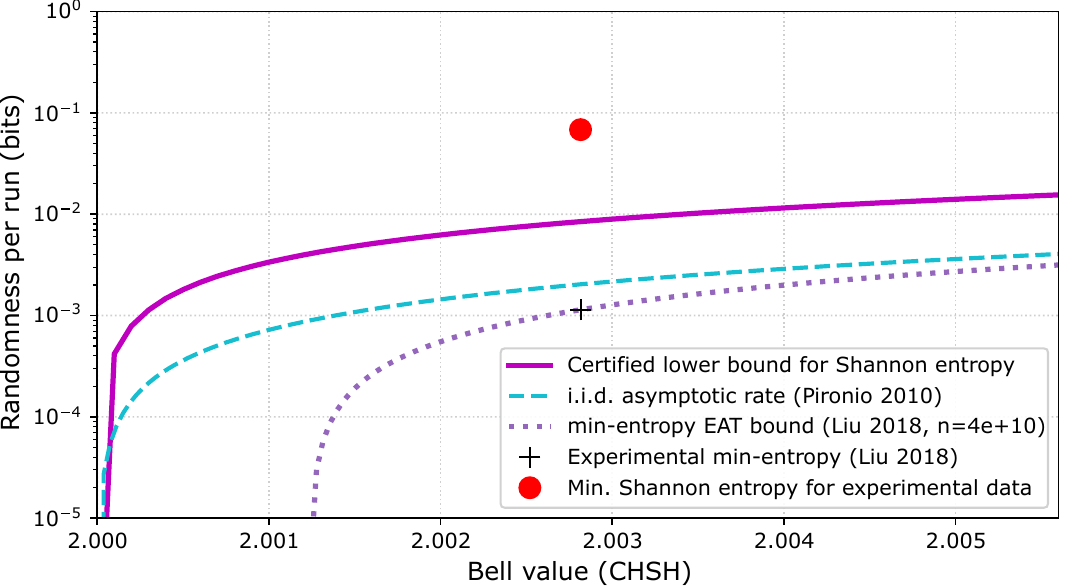}
	\caption{(color online) Comparison of extractable randomness certified by Shannon entropy and min-entropy benchmarks, showing the higher randomness yield of the Shannon approach for finite-size experimental data.}
	\label{fig:paper2}
\end{figure}

We present the results in Figure \ref{fig:paper2}. It shows the extractable randomness per run, in bits on a logarithmic scale, as a function of the observed CHSH Bell value. The figure compares the continuous lower bound for Shannon entropy with two min-entropy benchmarks: the asymptotic i.i.d.\ rate derived \cite{pironio_random_2010} and the finite-size EAT bound calculated for $n = 4 \cdot 10^{10}$ trials. The figure includes the direct experimental result from \cite{Liu2018}, which indicates the certified finite-size min-entropy yield of $1.14 \cdot 10^{-3}$ bits per run. We also present the certified minimum Shannon entropy obtained based on the same probability distribution. The difference between the two points shows the higher randomness yield obtained with Shannon entropy certification compared to EAT-based min-entropy extraction.

The two methods reflect a trade-off between yield and security. Shannon entropy certification estimates a higher amount of certifiable randomness because it averages unpredictability over long sequences and does not focus exclusively on the most probable outcome. In contrast, the EAT-based min-entropy method produces lower certified rates, such as the experimentally demonstrated $1.14 \cdot 10^{-3}$ bits/run, because it assumes an adversary may guess the most likely outcome and therefore heavily discounts the raw entropy to account for statistical uncertainty. 

In \cite{Leone2022}, the authors implement an SDI QRNG based on single-photon entanglement generated using attenuated laser pulses and linear optics. To mitigate temporal correlations induced by detector dead time and after-pulsing, the analysis incorporates a Markovian memory model. Experimental imperfections in polarization optics are quantified through numerical penalties $e_P$ and $e_I$, which constrain deviations between ideal and observed probability distributions. The first penalty is derived by calculating the maximum mathematical difference between ideal detection probabilities and the actual probabilities altered by polarization flaws in the setup's optical components, while $e_I$ is then derived by propagating these deviations through the CHSH equation to find the maximum resulting penalty on the final Bell violation.

Randomness is quantified using min-entropy, yielding conservative extraction rates determined by the assumed trust level of internal device parameters. In the least trusted model, only transmission coefficients $t$ are fixed, while visibility $v$, phase mismatch $\delta$, and waveplate angles $\pi$ remain adversarially unconstrained, resulting in large error bounds and low certified entropy. Progressively fixing the waveplate angles and phase mismatch reduces uncertainty and tightens the bounds, increasing the certified min-entropy to \(H_{\min}^{*}=0.301\) with a generation rate of \(52.7\) kHz in the fully characterized configuration.

When we apply our certification model to the empirical data, the method substantially increases certified randomness across all trust levels. In the general scenario, the certified Shannon entropy is approximately $0.124$ bits per event, corresponding to a generation rate of $21.7$ kHz compared to the original $4.4$ kHz min-entropy rate. In the fully trusted configuration, the certified entropy reaches approximately $0.70$ bits per event, increasing the generation rate from $52.7$ kHz to over $122$ kHz without modifying the physical implementation.

We present the full results in Table \ref{tab:trust_comparison}, which compares certified randomness rates obtained under different levels of trust using both min-entropy and Shannon entropy-based methods. The table shows that, for the same experimental data and physical implementation, replacing min-entropy with Shannon entropy significantly increases the extractable randomness across all scenarios.

\onecolumngrid

\begin{table}[h]
\centering
\begin{tabular}{|l|c|c|c|c|c|}
\hline
Level of Trust & Original $H_{\min}^{*}$ & Original rate & Shannon entropy ($e_P$) & Shannon entropy ($e_P + e_I$) & Shannon entropy-based rate \\
\hline
General & 0.0250 & 4.4 kHz & 0.9969 & 0.1240 & 21.7 kHz \\
Fixed $t$ & 0.0630 & 11.0 kHz & 0.9969 & 0.2543 & 44.5 kHz \\
Fixed $t, \pi$ & 0.2690 & 47.1 kHz & 0.9972 & 0.6596 & 115.5 kHz \\
Fixed $t, \pi, \delta$ & 0.3010 & 52.7 kHz & 1.0691 & 0.6996 & 122.5 kHz \\
\hline
\end{tabular}
\caption{Comparison of certified randomness rates for experimental method presented in \cite{Leone2022}}
\label{tab:trust_comparison}
\end{table}

\twocolumngrid

In \cite{Lin2023}, the authors showed a DI non-interactive zero-knowledge proof protocol implemented using a QRNG certified via a loophole-free CHSH Bell test. Randomness certification is performed within the Quantum Probability Estimation framework (a mathematical framework that utilizes quantum estimation factors to establish a strict lower bound on the smooth conditional min-entropy of the classical-quantum state model against a predefined success threshold). Applied to experimental data exhibiting a CHSH value of $2.004$, the method yields a conservative min-entropy rate of $3.98 \cdot 10^{-3}$ bits per trial, which is $3.14$ times less than the analytical lower bound of Shannon entropy, which certifies approximately $1.25 \cdot 10^{-2}$ bits/trial. 

The last reference \cite{Shen2018} implements a CW SPDC source for detection-loophole-free quantum randomness generation. The measurement window was optimized to produce a mean photon-pair number of $\mu \approx 0.32$ per round, resulting in a measured CHSH Bell value of $S = 2.01602$. Randomness certification was performed using a fully DI framework. Randomness quantification was based on the smooth conditional min-entropy, which upper bounds the adversary’s optimal guessing probability for the generated output string, assuming access to a purifying quantum side information system. Under this criterion, the reported CHSH Bell value $B = 2.01602$ corresponds to an asymptotic certified min-entropy rate of approximately $5.8 \cdot 10^{-3}$ bits/round.

We apply the analytical Shannon entropy lower bound to the same model across a continuous range of pump powers, as shown in Figure \ref{fig:paper2}. The certified entropy profile is governed by the Poissonian emission statistics of the source. At low brightness, vacuum emissions dominate the signal, keeping correlations below the nonlocality threshold. At high pump powers, higher-order multi-pair emissions, together with non-photon-number-resolving bucket detectors, reduce entanglement visibility and suppress the Bell violation. The optimal operating point occurs near $\mu \approx 0.32$, where the ratio of single-pair emissions to multi-pair noise is highest. At this point, the experimental data certify a Shannon entropy of approximately $H = 3.79 \cdot 10^{-2}$ bits/round.

\begin{figure}[htbp]
	\includegraphics[width=\linewidth]{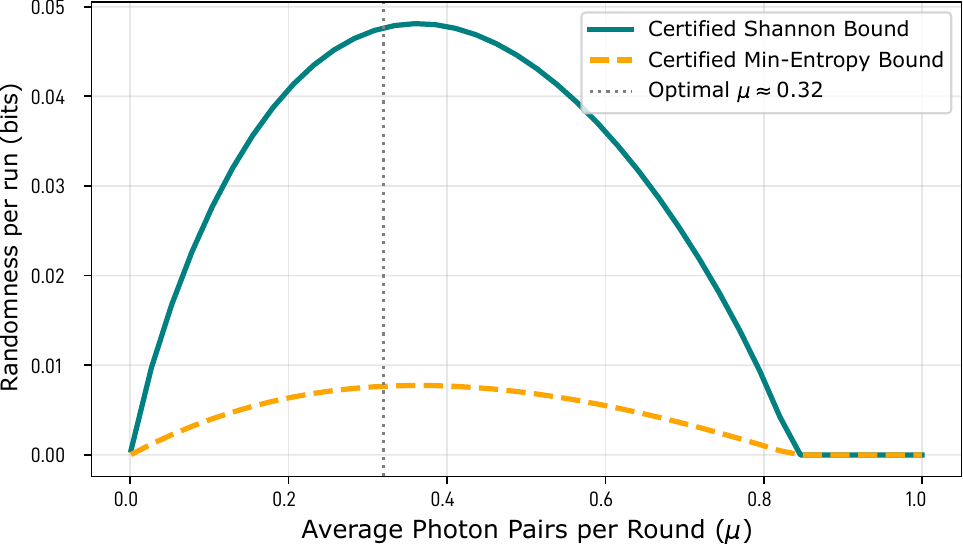}
	\caption{(color online) DI certified thoretical entropy bounds as a function of average photon pairs ($\mu$), illustrating that the analytical Shannon entropy lower bound certifies over six times more secure randomness than the traditional min-entropy bound at the experimental optimum of $\mu \approx 0.32$.}
	\label{fig:paper1-1}
\end{figure}

\section{Conclusions} \label{sec:conlusions}
In this work, we developed and applied a quantitative procedure for the systematic analysis of Bell expressions used in DI quantum randomness certification. The proposed approach significantly reduces computational overhead by utilizing entropy-based metrics derived from Shannon entropy bounding and approximation techniques~\cite{okuła2025deviceindependentshannonentropycertification}. This formulation enables large-scale numerical exploration of Bell inequalities and allows for efficient evaluation of their randomness-certifying potential. Using this method, we analyzed more than half a million Bell expressions corresponding to the $(4,3)$ measurement configuration and identified five particularly notable examples based on their entropy performance across varying levels of white noise. The comparison between analytical estimates and nonlinear optimization results revealed that the analytical entropy values act as upper bounds, underscoring the importance of nonlinear refinement for precise quantification of certifiable randomness. Beyond the entropy-based analysis, we extended our investigation to the self-testing properties of the identified Bell certificates using the self-testing for boxes formalism~\cite{banacki_hybrid_2022}. This complementary perspective enabled a deeper understanding of the underlying quantum correlations and their uniqueness under different noise conditions.

While DI protocols avoid assumptions about device internals, practical implementations are limited by noise and finite detection efficiency. These imperfections reduce the observed Bell violation $\langle\mathcal{B}\rangle$, lowering the certified entropy until the classical bound is reached. Detection loopholes can be addressed by assigning predetermined outcomes to no-click events, which imposes a minimum detection efficiency for positive randomness certification \cite{PhysRevA.47.R747}. In semi-DI frameworks, assumptions such as trusted state preparation or bounded Hilbert-space dimension relax the requirement for loophole-free Bell tests. Our entropy-bounding method takes these constraints into consideration, enabling randomness certification more realistic scenarios.

Our findings bridge a longstanding methodological gap between physical entropy sources and classical cryptographic standards. While classical cryptography relies on Shannon entropy to establish the theoretical limits of perfect secrecy and key-size lower bounds, the characterization of quantum random number generators has predominantly focused on min-entropy to mitigate worst-case hardware biases. By strictly certifying the Shannon entropy of a quantum source, we enable a direct and rigorous comparison between quantum methods and other randomness-generating options. This alignment enables us to more thoughtfully evaluate the advantages of quantum randomness, moving beyond isolated device-specific metrics to a unified framework that quantifies security against the standard of perfect unpredictability.

Furthermore, while we introduced the theoretical concept of Shannon entropy certification in \cite{okuła2025deviceindependentshannonentropycertification}, the present work demonstrates the utilization of this framework in a practical setting. The comprehensive search methodology and results derived herein facilitate the integration of these randomness extraction protocols into physical devices, directly enhancing their real-world applicability. Ultimately, these findings significantly enhance the quality and robustness of randomness generation by providing a concrete framework for the exploration of certification protocols that maintain high entropy rates even in the presence of unavoidable noise.

\section{Acknowledgements}
We acknowledge the use of a computational server financed by the Foundation for Polish Science (IRAP project, ICTQT, contract no. 2018/MAB/5/AS-1, co-financed by the EU within the Smart Growth Operational Programme). We also acknowledge the use of a computational server shared by the Immunoinformatics team (Alfaro J., Daghir-Wojtkowiak E., Kallor A., Pałkowski A., Waleron M.) of ICCVS, financed by the EC (Horizon2020 project KATY GA no. 101017453). PM was supported by the European Union’s Horizon Europe research and innovation programme under grant agreement No 101080086/NeQST.

NPA optimization was implemented using Python library ncpol2sdpa \cite{Wittek_2015}. We used MOSEK \cite{mosek} as a solver. nonlinear optimization was achieved using SciPy \cite{2020SciPy-NMeth}, using COBYQA as the optimization solver \cite{rago_thesis} \cite{cobyqa}.

\bibliographystyle{plainnat}
\bibliography{shannon}

\end{document}